\documentclass{aa}  
\usepackage{graphicx}
\usepackage{txfonts}
\usepackage{xcolor}
\usepackage{placeins} %for images
\usepackage{caption}
\usepackage{placeins} %for images
\usepackage{wrapfig}
\usepackage{siunitx}
\usepackage{amsmath}

\usepackage{natbib,twoopt}
\definecolor{links}{rgb}{0, 0, 255}
\usepackage[colorlinks=true,allcolors=links]{hyperref} %% to avoid \citeads line fills
\bibpunct{(}{)}{;}{a}{}{,}             %% natbib format for A&A and ApJ
\makeatletter
  \newcommandtwoopt{\citeads}[3][][]{\href{http://adsabs.harvard.edu/abs/#3}%
    {\def\hyper@linkstart##1##2{}%
     \let\hyper@linkend\@empty\citealp[#1][#2]{#3}}}
  \newcommandtwoopt{\citepads}[3][][]{\href{http://adsabs.harvard.edu/abs/#3}%
    {\def\hyper@linkstart##1##2{}%
     \let\hyper@linkend\@empty\citep[#1][#2]{#3}}}
  \newcommandtwoopt{\citetads}[3][][]{\href{http://adsabs.harvard.edu/abs/#3}%
    {\def\hyper@linkstart##1##2{}%
     \let\hyper@linkend\@empty\citet[#1][#2]{#3}}}
  \newcommandtwoopt{\citeyearads}[3][][]%
    {\href{http://adsabs.harvard.edu/abs/#3}
    {\def\hyper@linkstart##1##2{}%
     \let\hyper@linkend\@empty\citeyear[#1][#2]{#3}}}
\makeatother

\newcommand{\kms}{\mathrm{km\,s^{-1}}}
\newcommand{\msun}{M_{\sun}}

\newcommand{\hi}{H{~\sc i}\xspace}
\newcommand{\hii}{H{~\sc ii}\xspace}
\newcommand{\hj}{\textrm{H \textsc{i}}}

\begin{document} 

\title{The \hi mass function of the Local Universe: Combining measurements from HIPASS, ALFALFA, and FASHI}
\titlerunning{\hi Mass Function of the Local Universe}
\author{
Wenlin Ma\inst{\ref{inst1},\ref{inst2}}
\and Hong Guo\inst{\ref{inst1}\fnmsep\thanks{Corresponding Author; guohong@shao.ac.cn}}
\and Haojie Xu\inst{\ref{inst1}} 
\and Michael G. Jones\inst{\ref{inst3}}
\and Chuan-Peng Zhang\inst{\ref{inst4},\ref{inst5}\fnmsep\thanks{Corresponding Author; cpzhang@nao.cas.cn}}
\and Ming Zhu\inst{\ref{inst4},\ref{inst5}\fnmsep\thanks{Corresponding Author; mz@nao.cas.cn}}
\and Jing Wang\inst{\ref{inst6}}
\and Jie Wang\inst{\ref{inst4}}
\and Peng Jiang\inst{\ref{inst4},\ref{inst5}}
}

\institute{
Shanghai Astronomical Observatory, Chinese Academy of Sciences, Shanghai 200030, China.\label{inst1} 
\and University of Chinese Academy of Sciences, Beijing 100049, China.\label{inst2} 
\and Steward Observatory, University of Arizona, 933 N Cherry Ave., Tucson, AZ 85721, USA. \label{inst3}
\and National Astronomical Observatories, Chinese Academy of Sciences, Beijing 100101, China.\label{inst4} 
\and Guizhou Radio Astronomical Observatory, Guizhou University, Guiyang 550000, China.
\label{inst5}
\and Kavli Institute for Astronomy and Astrophysics, Peking University, Beijing 100871, China.\label{inst6}
}

\abstract
% 5 {} token are mandatory
  % context heading (optional)
  {We present the first \hi mass function (HIMF) measurement for the recent FAST All Sky \hi (FASHI) survey and the most complete measurements of the HIMF in the Local Universe thus far. We obtained these results by combining the \hi catalogues from \hi Parkes All Sky Survey (HIPASS), Arecibo Legacy Fast ALFA (ALFALFA), and FASHI surveys at a redshift of $0<z<0.05$, covering 76\% of the entire sky. We adopted the same methods to estimate the distances, calculate the sample completeness, and determine the HIMF for all three surveys. The best-fit Schechter function for the total HIMF shows a low-mass slope parameter of $\alpha = -1.30 \pm 0.01$ and a `knee' mass of $\log (M_s/h_{70}^{-2}\msun) = 9.86 \pm 0.01,$ along with a normalisation of $\phi_s = (6.58 \pm 0.23) \times 10^{-3}\,h_{70}^3\,{\rm Mpc ^{-3} dex^{-1}}$. This gives us the cosmic \hi abundance: $\Omega_{\hj} = (4.54 \pm 0.20)\times 10^{-4}\,h_{70}^{-1}$. We find that a double Schechter function with the same slope  $\alpha$ better describes our HIMF, where the two different `knee' masses are $\log(M_{s_1}/h_{70}^{-2}\msun) = 9.96 \pm 0.03$ and $\log(M_{s_2}/h_{70}^{-2}\msun) = 9.65 \pm 0.07$. We verify that the measured HIMF is marginally affected by the choice of distance estimates. The effect of cosmic variance is significantly suppressed by combining the three surveys and this provides a unique opportunity to obtain an unbiased estimate of the HIMF in the Local Universe. }
    % conclusions heading (optional)

   
\maketitle
        
\section{Introduction}
Neutral hydrogen, in its atomic (\hi) and molecular (H$_2$) forms, plays an important role in the galaxy baryon cycle \citep[see e.g.][for reviews]{Peroux2020, Saintonge2022}. Although H$_2$ serves as  direct fuel for star formation, \hi serves as a reservoir to forming H$_2$. Understanding the distribution of \hi and how it is correlated with the properties of galaxies is crucial for theoretical studies of galaxy formation and evolution.

The two most important measurements for describing the \hi content are the cosmic \hi abundance ($\Omega_\hj$) and the \hi mass function (HIMF). In brief, $\Omega_{\hj}$ quantifies the total \hi mass in the universe and its evolution with redshift, while $\Omega_\hj(z)$, is closely related to the star formation history of galaxies \citep[e.g.][]{Rafieferantsoa2019, Kamphuis2022}. As the counterpart of the galaxy stellar mass function in an optical survey, the HIMF describes the number densities of galaxies in different \hi mass bins and provides the mass distribution of the \hi gas in addition to the total abundance of $\Omega_\hj$. In the Local Universe, $\Omega_{\hj}$ can be accurately determined by directly summing up the HIMF. At higher redshifts, $\Omega_\hj$ is usually estimated using the stacked \hi signals and  the damped Ly$\alpha$ systems, albeit with large uncertainties \citep[see][and references therein]{Peroux2020}. 

The HIMF is not only useful for deriving $\Omega_\hj$, but it also encodes essential information about galaxy assembly histories. Since the \hi gas distribution is very sensitive to accretion and feedback mechanisms \citep[e.g.][]{Fu2013,Popping2015,Xie2017,Guo2022}, the HIMF serves as a valuable tool to distinguish between various galaxy formation models, where the galaxy stellar mass functions at low redshifts are typically well reproduced \citep[e.g.][]{Baugh2019,Dave2020}. The HIMF also shows a strong dependence on the halo and large-scale environment \citep[e.g.][]{Zwaan2005,Jones2020,Ma2024}. Precise measurements of the HIMF are also the key science goal of current and future \hi surveys, including  Widefield ASKAP L-band Legacy All-sky Blind Survey \cite[WALLABY;][]{Koribalski2020},  MeerKAT International GigaHertz Tiered Extragalactic Exploration \citep[MIGHTEE;][]{Jarvis2016}, and  Five-hundred-metre Aperture Spherical radio Telescope (FAST) All Sky \hi survey \cite[FASHI;][]{Zhang2024}.

In the Local Universe ($z<0.06$), the HIMF has been directly measured by the \hi Parkes All-Sky Survey \citep[HIPASS;][]{Barnes2001, Meyer2004} and the Arecibo Legacy Fast ALFA Survey \cite[ALFALFA;][]{Giovanelli2005, Haynes2011}. It has been found that the measured HIMF, $\phi(M_{\hj})$, can be well described by a Schechter function \citep{Schechter1976}, 
\begin{equation}
    \phi(M_{\hj}) = \frac{dn}{d \log M_{\hj}} = \ln(10) \phi_s \left(\frac{M_\hj}{M_s}\right)^{\alpha+1} \exp\left(-\frac{M_\hj}{M_s}\right),\label{eq:HIMF}
\end{equation}
where $\phi_s$ is the normalization, $\alpha+1$ is the low-mass end slope, and $M_s$ is the `knee' mass. 
 
\cite{Zwaan2005} used HIPASS and made one of the first HIMF measurements at $z\sim0$, finding a `knee' mass of $\log(M_s/h_{70}^{-2}\msun)=9.80\pm0.03$ and a slope of $\alpha = -1.37\pm0.03$. Using the 40\% complete sample of ALFALFA that has much better sensitivity and resolution than HIPASS, \cite{Martin2010} found a higher `knee' mass ($\log(M_s/h_{70}^{-2}\msun)=9.96\pm0.02$) and a slightly flatter slope ($\alpha = -1.33\pm0.03$). The HIMF measurement was later updated by \cite{Jones2018} with the final ALFALFA sample and they found a consistent `knee' mass of $\log(M_s/h_{70}^{-2}\msun) = 9.94$, but a much shallower slope ($\alpha = -1.25\pm0.02$). 

In optical surveys, galaxy stellar mass estimates may depend on various assumptions of initial mass functions, dust extinction laws, and stellar population synthesis models. The \hi mass estimate of a target galaxy has the great advantage that it suffers  much less from systematics and it mainly depends on the uncertainties in the source distance ($D_L$) and the integrated \hi line flux ($S_{21}$; \citealt{Meyer2017}), as well as the unquantified self-absorption line. However, most blind \hi surveys are not volume-limited in nature. The number of measured \hi targets depends both on $S_{21}$ and the width of the line profile ($W_{50}$). Galaxies with higher flux and narrower line profiles are much easier to detect \citep{Giovanelli2005,Haynes2011}. Therefore, it is crucial to quantify the sample completeness due to the selection effect. As shown in \cite{Guo2023}, the differences between the HIMF measurements of \cite{Martin2010} and \cite{Jones2018} are caused by both cosmic variance and the adopted completeness cuts. Since the 50\% completeness cut is used in \cite{Jones2018} to derive the HIMF,  the number densities of low-$M_\hj$ galaxies between 50\% and 100\% completeness cuts are underestimated, leading to the shallower low-mass slope.

As investigated in \cite{Jones2018}, the systematic uncertainties in the HIMF caused by distance estimates are very minor, namely, only altering values of $\log M_s$ and $\alpha$ on the level of $0.01$. The remaining source of systematic uncertainties, and probably the most important one when comparing the HIMFs of different surveys, is the cosmic variance effect. That is, the intrinsic HIMFs in different survey volumes, no matter how accurately they are measured, could vary from each other. This effect was already seen in the HIMF measurements reported for the final sample of ALFALFA. The differences between the separate measurements of $\alpha$ in the spring and fall sky regions of ALFALFA are 0.14, even when all other conditions are the same \citep[see Fig.~3 of][]{Jones2018}. The apparent discrepancies between the HIMFs of HIPASS (southern sky) and ALFALFA (northern sky) could be caused by variations in galaxy populations and large-scale structures \citep{Ma2024}, as well as the different methods of estimating sample completeness and target distances.

However, galaxies with low \hi masses can only be probed in a very limited redshift range, since the \hi flux for a given \hi mass would decrease rapidly as the distance increases ($S_{21}\propto D_L^{-2}$). To minimise the cosmic variance effect, we can increase the volume by conducting deeper \hi surveys or covering a larger sky area. Recently, the first catalogue of the FASHI survey was  released \citep{Zhang2024}. It features significantly improved sensitivity, resolution, and depth, compared to previous surveys \citep[see e.g.][]{Wang2022}. This catalogue covers approximately 7600 square degrees of the sky, which is also complementary to the existing HIPASS and ALFALFA sky coverage. It offers an unprecedented opportunity to measure the most accurate HIMF in the Local Volume by combining the three survey catalogues. 

In this paper, our aim is to measure the HIMF by combining the \hi sources in the HIPASS, ALAFLFA, and FASHI surveys, with a total sky coverage of around 31,528~$\deg^2$ (i.e. nearly 76\% of the entire sky). Most importantly, we  aim to process all three catalogues using the same set of distance estimates, sample completeness corrections, and the HIMF calculation method. The resulting HIMF will provide an important reference for future \hi surveys.

The organisation of this paper is as follows. In Sect.~\ref{sec:data}, we introduce the observational data used in the constraints. Section~\ref{sec:method} describes the methods that we use in estimating HIMF. We show the results in Sect.~\ref{sec:results}. The discussion and conclusions are presented in Sects.~\ref{sec:discussion} and~\ref{sec:conclusion}, respectively. Throughout the paper, all masses are expressed in units of $\msun$. We adopted a flat Lambda cold dark-matter cosmology of $\Omega_{\rm m}=0.3$ and the Hubble constant is assumed to be $H_0=70\,h_{70}\,{\rm km\,s^{-1}\,Mpc^{-1}}$.

\section{Data}\label{sec:data}

\begin{figure*}
    \centering
    \includegraphics[width=\textwidth]{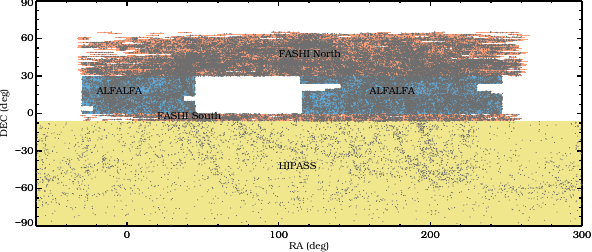}
    \caption{Angular distribution of \hi sources in FASHI sky (orange area), ALFALFA sky (blue area), and HIPASS (yellow area) sky. Grey points indicate individual detections. As the sky coverage of FASHI survey is not uniform, we split it into several pixels, each with an area of 2 $\rm deg^2$. In cases where there is an overlap between two surveys, we present the areas with deeper coverage.}
    \label{fig:radec}
\end{figure*}

In this work, we combined the HIPASS, ALFALFA, and FASHI catalogues to measure the local HIMF at $0<z<0.05$. In Fig.~\ref{fig:radec} we show the angular distributions of \hi sources in FASHI (grey points in red regions), ALFALFA (grey points in blue regions), and HIPASS (grey points in yellow regions). In cases of overlaps between two surveys, we only adopted the data from a given survey with a higher sensitivity. Table~\ref{tab:survey} displays the details of the samples in three surveys, including geometric cuts in the declination (Dec.). We note that the sample size contains only galaxies above the 50\% completeness limits (Sect.~\ref{subsec:complete}). The ranges of declination are determined to avoid overlap between the surveys.

\begin{table*}
\setlength{\tabcolsep}{5pt} 
\renewcommand{\arraystretch}{1.3} 
\caption{Details of samples we adopted in three surveys, including redshift ranges, sample sizes, Dec. ranges, and sky areas.}\label{tab:survey}
\centering
\begin{tabular}{ c c c c c} 
 \hline
 \hline
Survey & $z$ & Sample size  & Dec. range & Sky area ($\rm deg^2$) \\ 
 \hline
 HIPASS & $0<z<0.042$ & 3079 & $\rm -90\si{\degree} < Dec. < -6.5\si{\degree} $ & 18291  \\ 
 ALFALFA & $0<z<0.05$ & 18713 & $\rm 0\si{\degree} < Dec. < 30\si{\degree}$ & 5649  \\ 
 FASHI north & $0<z<0.05$ & 16879 &  $\rm 30\si{\degree} < Dec. < 66\si{\degree}$  & 5942  \\ 
 FASHI south & $0<z<0.05$ & 2720 &  $\rm -6.2\si{\degree} < Dec. < 0\si{\degree}$  & 1646  \\ 
 \hline
\end{tabular}
\end{table*}

\subsection{HIPASS}
HIPASS was carried out with the Parkes 64~m radio telescope in Australia. It is the first blind \hi survey to cover the entire southern sky with the declination ranging from $-90^\circ$ to $+2^\circ$ \citep{Meyer2004}. The HIPASS survey also has a northern extension catalogue covering the range of $+2^\circ<{\rm Dec} < +25.5^\circ$ \citep{Wong2006}, but it was not included in the HIMF calculation \citep{Zwaan2005}. It provides the largest uniform \hi catalogue in the southern sky, with 4315 sources in the redshift range of $-1280\,\kms<cz_{\odot}<12700\,\kms$. The Parkes beam diameter is 15.5~arcmin at 21~cm, and the root-mean-square (rms) noise of HIPASS is 13.3 mJy $\rm beam ^{-1}$ at a velocity resolution of $26.4\,\kms$. As a first-generation survey, the mean depth of HIPASS was relatively shallow and it also suffers from a particularly large beam size (which can lead to source confusion and blending). In this paper, we limit the HIPASS sample to the declination range of $-90^\circ<{\rm Dec.}<6.5^\circ$ to avoid overlap with the FASHI south data.

\subsection{ALFALFA}
ALFALFA is a second-generation \hi survey using the 305~m Arecibo single-dish telescope. Compared to HIPASS, it is evident that ALFALFA has a much smaller beam diameter (3.8~arcmin at 21~cm) and the rms noise level is significantly improved (2.4~mJy per beam at a velocity resolution of $10\,\kms$). The final data release includes $\sim 31500$ extragalactic sources and covers almost $6900\,\deg^2$ of the northern sky in the redshift range of $-2000\,\kms<cz_{\odot}<18000\,\kms$ \citep{Haynes2018}. The ALFALFA footprint is split into two continuous regions, which are named the `spring sky' with right ascension (RA) ranges in $\rm 07^h 30^m < RA < 16^h 30^m$ and `fall sky' ranges in $\rm 22^h < RA < 03^h$ according to their observation seasons. The declination ranges from $0^\circ$ to $36^\circ$ in both regions. A drift scan strategy was employed to observe both regions, resulting in high time efficiency and uniform coverage. In this paper, we only include the Code 1 sources (signal-to-noise ratio larger than 6.5) in the ALFALFA catalogue, as the reliability is close to 100\% \citep{Saintonge2007}. We follow the boundary cuts as in \cite{Jones2018} and limit the declination range to less than $30^\circ$ to avoid overlap with FASHI. To be consistent with the HIMF measurement of \cite{Jones2018}, we also limit the redshift range of ALFALFA to $0<z<0.05$, beyond which radio frequency interference (RFI) becomes more severe. 

\subsection{FASHI}\label{subsec:data_fashi}
Based on the FAST 500-m single-dish radio telescope, FASHI has achieved a greater survey depth compared to both HIPASS and ALFALFA, as well as the smaller beam size (effectively $3.24$ arcmin at 21~cm; \citealt{Wang2023,Wang2024}) and much lower rms noise level ($1.5$~mJy per beam at $6.4\,\kms$ resolution; \citealt{Zhang2024}). FASHI was designed to observe the entire detectable sky of FAST in the declination range of $\rm -14^\circ< Dec < +66^\circ$ (around $22000\,\deg^2$). The first data release \citep{Zhang2024} covers two separate regions, which we refer to as `FASHI north' ($30^\circ < {\rm Dec.} < 66^\circ$) and `FASHI south' ($-6.2^\circ < {\rm Dec.} < 0^\circ$), both with the right ascension in the ranges of $\rm 0^h \leq RA \leq 17.3^h$ and $\rm 22^h \leq RA \leq 24^h$. The observed redshift range of FASHI is $200\,\kms< cz_{\odot}< 26323\,\kms$ with a frequency range of 1305.5--1419.5~MHz. However, this frequency range includes radio recombination lines, which are produced by gas ionised by young massive stars within \hii regions of the Milky Way. To identify and eliminate these lines from the FASHI data, they used the criteria to ensure that the same spatial location exhibits consistent flux density and line width across various transition frequencies. In total, 41741 extragalactic sources have been detected. For fair comparisons with ALFALFA, we adopt the same redshift limits as $0<z<0.05$. Although FASHI observed quite a few galaxies in $0.05<z<0.09$, the influence of RFI there would be much more significant for our HIMF measurements. 

Unlike ALFALFA, FASHI is carried out in a time-filler mode, that is, the observations are made when there were no other running programmes in the observing queue. This observation strategy made full use of the available time to increase the sample size, but results in an inhomogeneous survey depth. The sky areas sampled multiple times would be much deeper than in other regions. As shown in Fig.~5 of \cite{Zhang2024}, the detection rms noise (in units of mJy per beam) varied strongly in different parts of the sky, as well as between FASHI north and FASHI south. FASHI south has a significantly higher detection rms noise (i.e. much lower source surface densities). Therefore, in this study, we decided to treat FASHI north and FASHI south as two separate samples. 

The problem of inhomogeneous sky coverage of FASHI is  similar to that of angular variations of observed galaxy surface number densities due to the foreground stars in optical surveys. Therefore, we followed the strategy of \cite{Xu2023} by applying correction weights to galaxies in different areas. To do so, we split the FASHI sky coverage into grid pixels of equal area of $2\,\deg^2$, which roughly matches the typical scale of the rms noise variation. Since the FASHI drift scans were performed at a fixed sky declination, we constructed the pixels in linear declination bins of $\Delta{\rm Dec.}=0.5^\circ$ and adjusted the number of RA bins in different Dec. bins to reach the same pixel area. The resulting sky coverage of FASHI north and FASHI south is shown as the red regions in Fig.~\ref{fig:radec}. The distribution at the edge of the survey is discrete, but will be much improved in the future FASHI data release.

\begin{figure*}
\centering
\includegraphics[width=0.8\columnwidth]{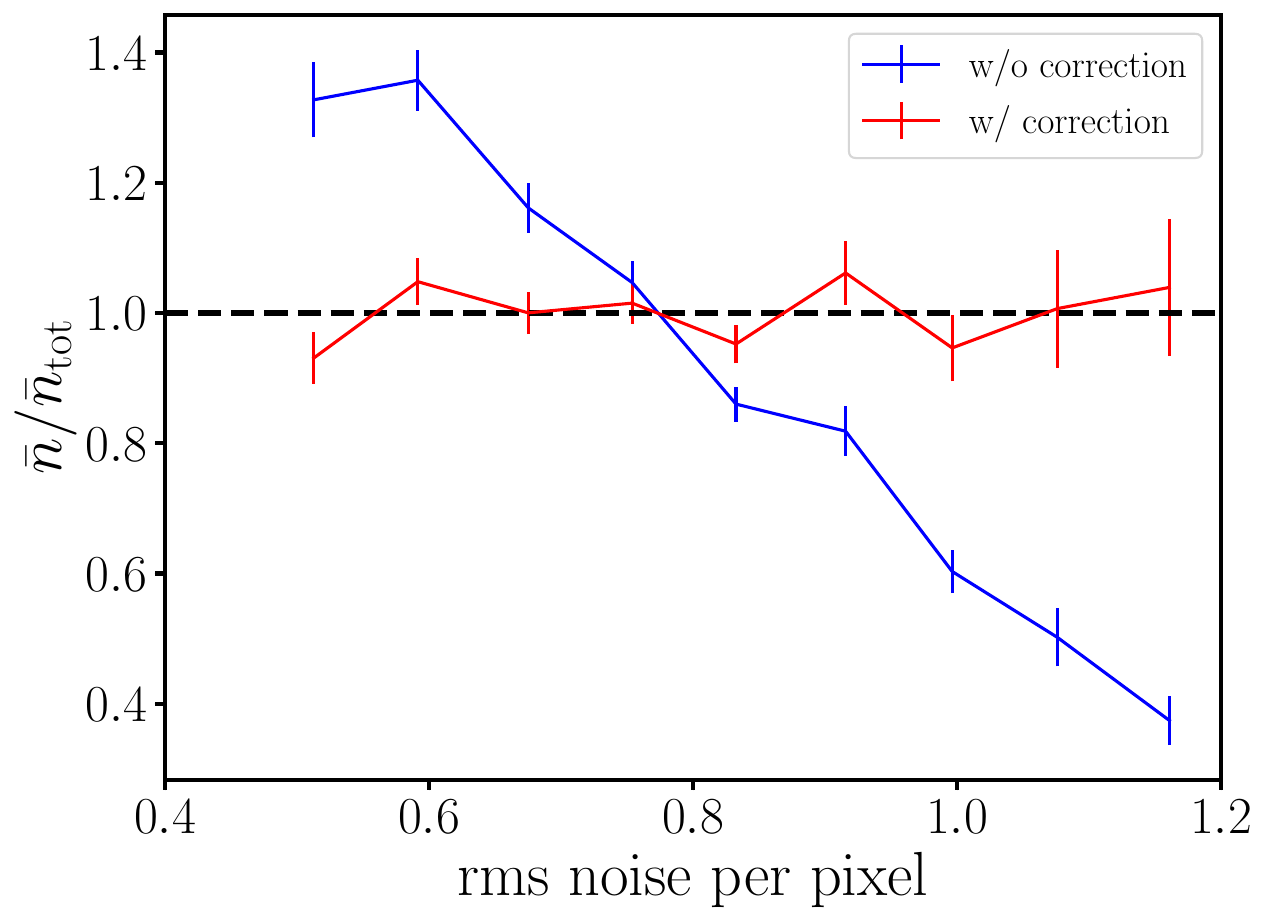}\includegraphics[width=0.8\columnwidth]{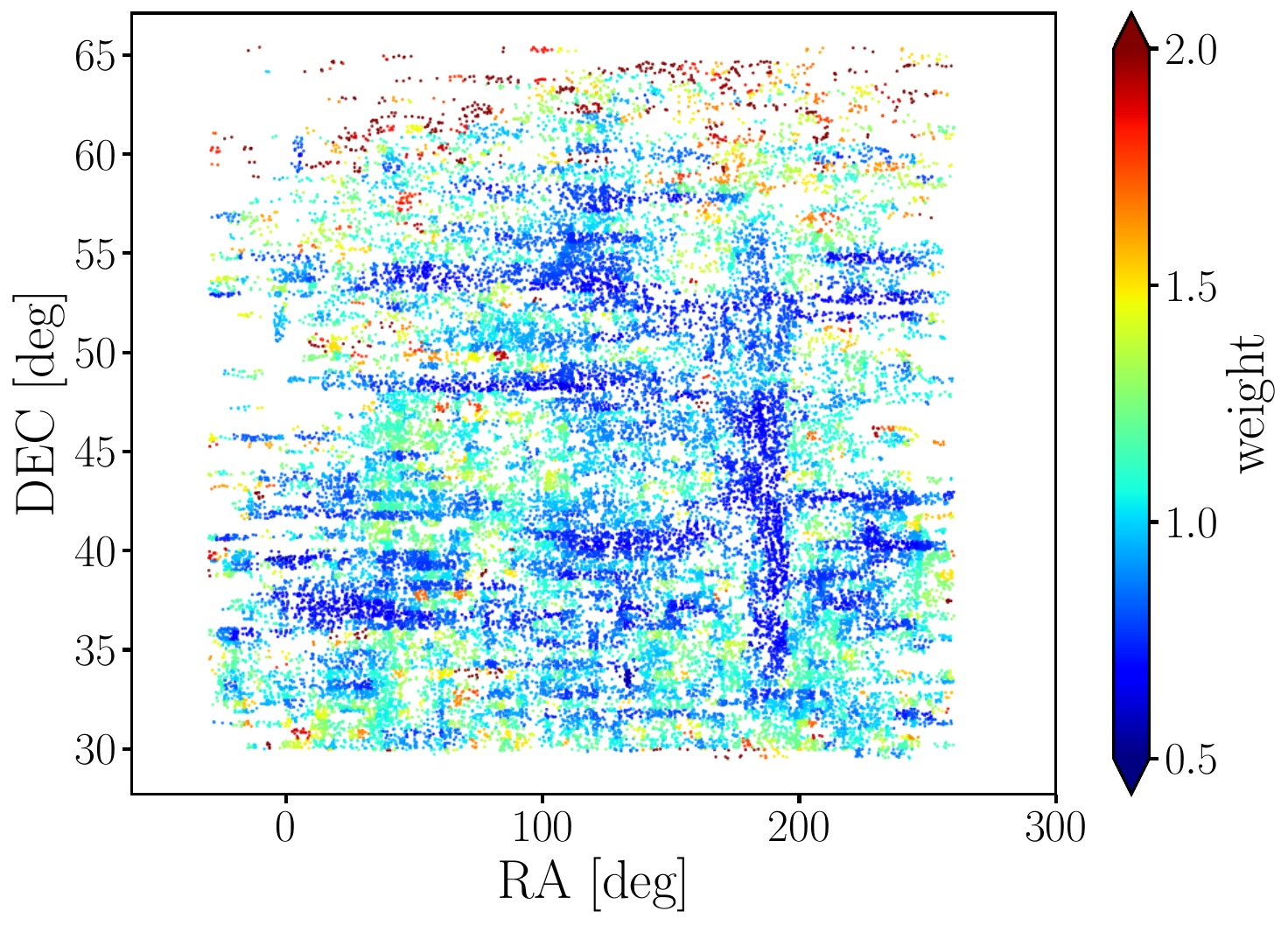}
\caption{Weights of FASHI north galaxies. Left panel: Blue line shows the normalised surface density as a function of the rms noise of the FASHI galaxies in each pixel. We weighted each galaxy with the reciprocal of the best-fitting rms noise dependence function to correct the surface density to be homogeneous, shown as red line. Right panel: Weights in different pixels for FASHI north. We only used pixels with the weights in the range of 0.5--2 to prevent  heavily weighted galaxies ending up assigned to less-sampled regions.}
\label{fig:weight}
\end{figure*}

Ideally, for a homogeneous \hi survey, there should be no strong spatial variation in the surface number densities of galaxies. A correction factor of the galaxy surface density in each pixel is needed to derive the correct HIMF for FASHI.  To find it, we first calculated the galaxy surface number density ($\bar{n}$) and the median rms noise for each pixel. In the left panel of Fig.~\ref{fig:weight}, we show the normalised surface number density, $\bar{n}/\bar{n}_{\rm total}$, as a function of the median rms noise of each pixel (blue line with errors), where $\bar{n}_{\rm total}$ is the average surface number density of the entire sample. It is evident that without any correction, the detected number of galaxies in each pixel is highly correlated with the rms noise level. More galaxies tend to be found in regions that have been scanned multiple times (with lower rms noise). To correct for this effect, we adopt a second-order polynomial function to fit the rms noise dependence and weight galaxy by the reciprocal of the best-fitting function. We note that the assignment of the correction factor is pixel-wise (i.e. all galaxies in the same pixel share the same weight). Galaxies in the high (low) surface density regions are therefore down-weighted (up-weighted). After applying the correction factor (denoted as $f_{\rm rms}$ and applied in the following calculation of the HIMF in Eq.~\ref{eq:phi}), the weighted source surface density is almost homogeneous (red line). The weights in different pixels for FASHI north are shown in the right panel of Fig.~\ref{fig:weight}. Most of the weights are in the range of 0.5--2 and we discarded all pixels with weights larger than 2 to refrain from heavily weighting galaxies in less-sampled regions. The weights for FASHI north and FASHI south were calculated independently, based on their own galaxy samples.

\section{Methods}\label{sec:method}
The \hi mass of each galaxy is determined from the flux following the standard equation \citep{Meyer2017}: 
\begin{equation}
    \frac{M_{\hj}}{\msun} = \frac{2.356 \times 10^5}{1+z} \left(\frac{D_L(z)}{\rm Mpc}\right)^2 \frac{S_{21}}{\rm Jy \ km \ s^{\rm -1}},
\end{equation}
where $S_{\rm 21}$ is the integrated \hi flux in units of Jy\,$\kms$ and $D_L(z)$ is the luminosity distance to the galaxy in units of Mpc. In order to minimise systematic uncertainties, we recalculated the \hi mass in three catalogues in the same manner. 

\subsection{Distance estimates} \label{subsec:cal_dis}
One of the main systematics of determining $M_\hj$ is the uncertainty of $D_L(z)$. Pure Hubble flow distances are commonly used for $cz_\odot>6000\,\kms$ \citep{Zwaan2005,Haynes2011}, where the contribution of the peculiar velocity is relatively small. However, different flow models adopted to estimate peculiar velocities at $cz_\odot<6000\,\kms$ could potentially cause large systematic uncertainties in $D_L(z)$ \citep{Masters2004}. \cite{Masters2005} proposed a Local Volume flow model to reduce the distance errors by considering the infall of galaxies onto local superclusters. However, the distinct bias related to the line-of-sight galaxy density distribution, known as the Malmquist bias \citep{Lynden1988}, will lead to incorrect assignment of peculiar velocities \citep{Strauss1995}. 

The Cosmicflows-4 Distance–Velocity Calculator \citep{Kourkchi2020}\footnote{https://edd.ifa.hawaii.edu/CF4calculator/} is designed to mitigate Malmquist bias and the asymmetry in velocity errors in translations to distance from the logarithmic modulus. It is based on the Cosmicflows-4 catalogue \citep{Tully2023} and consists of two calculators. One is based on the smoothed velocity field from the numerical action methods \citep[NAM;][]{Shaya2017} model, but is limited only to a distance of 38 Mpc. The other CF4 calculator is based on the Wiener filter model \citep{Valade2024} and extends to 500 Mpc. 

The original HIPASS catalogue used the pure Hubble flow distances \citep{Meyer2004,Zwaan2005} and ALFALFA adopted a local volume flow model of \cite{Masters2005} for $cz_\odot<6000\,\kms$. FASHI used the NAM model of Cosmicflows-3 Distance–Velocity Calculator for $cz_\odot<2400\,\kms$ and the CF3 model \citep{Graziani2019} for $2400\,\kms<cz_\odot<15000\,\kms$ and the pure Hubble flow for $cz_\odot>15000\,\kms$. In this work, we applied the updated Cosmicflows-4 model of \citep{Valade2024} to all three catalogues. As shown in Sect.~\ref{subsec:distance_error}, the influence of different flow models on the HIMF measurements in this study is minor. 

\subsection{Sample completeness}\label{subsec:complete}
The completeness of the \hi sample is the main uncertainty when estimating the HIMF. One way to estimate completeness is to insert a large number of synthetic sources into the data. The completeness can then be determined from the rate of recovered synthetic sources \citep{Rosenberg2002, Zwaan2004}. In this study, we followed the method of \cite{Haynes2011}, using concrete data  to calculate the completeness limits for three surveys. Since completeness is a function of both $S_{21}$ and $W_{50}$, we divided $\log W_{50}$ into 20 bins from 1.0 to 3.0 with a bin width of 0.1. In each $\log W_{50}$ bin, we calculated the surface number density of galaxies in logarithmic intervals of $S_{21}$, $dn/d\log S_{21}$, as a function of $\log S_{21}$. The surface density, rather than the number of galaxies in each $\log S_{21}$ bin as in Fig.~11 of \cite{Haynes2011}, is used to make fair comparisons among samples from different survey areas. 

\begin{figure*}
\centering
\includegraphics[width=0.9\textwidth]{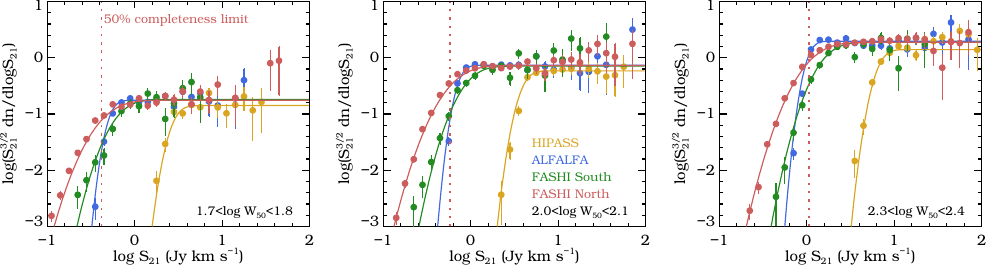}
\caption{Distribution of $S_{21}^{3/2}dn/d\log S_{21}$ as a function of the flux density, $S_{21}$, where $dn$ is the surface number density of galaxies in a given $\log S_{21}$ bin. This relation is used to determine the completeness of \hi targets for all three surveys in different $W_{50}$ bins, following the practice of \cite{Haynes2011}. For fair comparisons, the measurements in both ALFALFA and FASHI are limited to the redshift range of $0<z<0.05$. However, HIPASS only has  samples in $0<z<0.042$. The orange, blue, green, and red circles are shown for the HIPASS, ALFALFA, FASHI south, and FASHI north galaxies, respectively. The solid lines in each panel are the best-fit error function. The vertical red dotted lines indicate the 50\% completeness limit of FASHI north in each $W_{50}$ bin.}
\label{fig:s21}
\end{figure*}

We show the measurements of $S_{21}^{3/2} dn/d\log S_{21}$ in three representative $\log W_{50}$ bins in Fig.~\ref{fig:s21}. As discussed in \cite{Haynes2011}, the galaxy surface density, $dn/d\log S_{21}$, would be proportional to $S_{21}^{-3/2}$ for a complete sample. The deviation from a constant value of $S_{21}^{3/2} dn/d\log S_{21}$ marks the start of incompleteness. Following \cite{Haynes2011}, we can use an error function to fit the completeness $C(S_{21}|W_{50})$ in each $W_{50}$ bin, 
\begin{equation}
    C(S_{21}|W_{50})=\frac{1}{2}\left[1+\mathrm{erf}\left(\frac{\log S_{21}-\log S_{21,50\%}}{\sigma_{\log S_{21}}}\right)\right]
,\end{equation}
where the two free parameters are $S_{21,50\%}$ and $\sigma_{\log S_{21}}$. Here, $S_{21,50\%}$ is the 50\% completeness limit in a given $W_{50}$ bin (i.e. $C(S_{21}|W_{50})=0.5$ when $S_{21}=S_{21,50\%}$) and $\sigma_{\log S_{21}}$ characterises the slope of decreasing completeness at the low-$S_{21}$ end. In addition, we used a free parameter $A_p$ to fit the plateau value of $S_{21}^{3/2} dn/d\log S_{21}$ in each $W_{50}$ bin. The best-fitting curves are shown as solid lines in Fig.~\ref{fig:s21}. The 50\% completeness limit $S_{21,50\%}$ in each $W_{50}$ bin of FASHI north is shown as the vertical dotted line. 

It is remarkable that ALFALFA, FASHI north, and FASHI south share the same plateau value of $S_{21}^{3/2} dn/d\log S_{21}$ in each $W_{50}$ bin, which demonstrates the reliability of the FASHI measurements. The HIPASS sample has consistently lower plateaus than other surveys because it is limited to a smaller volume ($z<0.042$ rather than $z<0.05$), which leads to  lower surface densities. It is clear from the comparisons that FASHI north has the lowest rms noise, reaching a low flux density of $S_{21}\sim0.1 {\rm Jy\,\kms}$, which is about 0.5~dex deeper than ALFALFA. HIPASS is considerably shallower than other surveys and only detects high $S_{21}$ sources. ALFALFA and FASHI south have comparable survey depths, which validates our separate treatments of FASHI north and FASHI south. 

\begin{figure}
\centering
\includegraphics[width=\columnwidth]{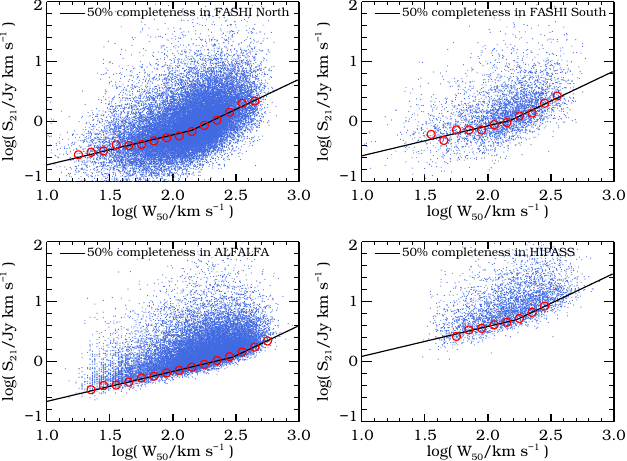}
\caption{Distribution of FASHI north (top-left), FASHI south (top-right), ALFALFA (bottom-left), and HIPASS (bottom-right) galaxies in the $\log S_{21}$--$\log W_{50}$ plane. The blue dots are all galaxy samples in each survey. The red open circles are the $\log S_{21}$ limit of 50\% completeness in each $\log W_{50}$ bin. The black solid lines are the fitted broken power law relations. } 
\label{fig:w50s21}
\end{figure}

In Fig.~\ref{fig:w50s21}, we show the dependence of $S_{21,50\%}$ on $W_{50}$ as red circles for the four \hi samples. The galaxy distribution in each sample is shown as the blue points. It is interesting to note that most galaxies in ALFALFA and HIPASS are above the 50\% completeness limits. However, for FASHI north and FASHI south, there is still a large fraction of galaxies below the limits. This is related to the shallow slopes, $\sigma_{\log S_{21}}$, seen in Fig.~\ref{fig:s21}. 
%(also listed in Table~\ref{tab:s21_w50}). 
In Table~\ref{tab:s21_w50}, we list the numbers of galaxies above and below the 50\% completeness cuts in the four samples. The fraction of galaxies below the cuts are about 17\%, 9\%, 45\%, and 34\% for HIPASS, ALFALFA, FASHI north, and FASHI south, respectively.
Although FASHI is able to detect galaxies with low $S_{21}$, they are very incomplete. To avoid large corrections to low-completeness galaxies, we only used galaxies with $C(S_{21}|W_{50})>0.5$ to measure the HIMF. Currently, the fraction of discarded galaxies below the limits is quite large for FASHI, but it will be improved in the future with better sampling rates.

As in \cite{Haynes2011}, we also find that the function of $S_{21,50\%}(W_{50})$ can be well fitted with a broken power law, as follows,
\begin{equation}
    \log S_{21,50\%} =
    \begin{cases}
        0.5\log W_{50} - a_1 & \text{$\log W_{50} < W_{\rm cut}$}, \\
        \log W_{50} - a_2    & \text{$\log W_{50} \geq W_{\rm cut}$},
    \end{cases} 
    \label{eq:s21_w50}
\end{equation}
where $a_1$ and $a_2$ are the free parameters. The transition $W_{50}$ value, $W_{\rm cut}$, is simply $2(a_2-a_1)$. The best-fitting relations of the 50\% completeness limits are shown as solid lines in Fig.~\ref{fig:w50s21} and the parameters are listed in Table~\ref{tab:s21_w50}. 
Our derived values of $a_1$ and $a_2$ for ALFALFA differ slightly from those of \citet[namely:\ $a_1=1.207$ and $a_2=2.457$]{Haynes2011}. This is   because their measurements were based on the early sample of 40\% ALFALFA. Our values align well with those reported in \citet[$a_1 = 1.170$ and $a_2=2.420$]{Oman2022}, which are also based on the ALFALFA 100\% sample.

\begin{table*}
\setlength{\tabcolsep}{3pt} 
\renewcommand{\arraystretch}{1.3} 
\caption{Parameters in Eq.~\ref{eq:s21_w50} and the number of galaxies above and below 50\% completeness in the four samples.}\label{tab:s21_w50}
\centering
\begin{tabular}{ c c c c c c c } 
 \hline
 \hline
Survey & $a_1$ & $a_2$ & $W_{\rm cut}$ & $\sigma_{\log S_{21}}$ & $S_{21}\geq S_{21,50\%}$ & $S_{21}<S_{21,50\%}$ \\ 
 \hline
 HIPASS & 0.412 & 1.528 & 2.232 & 0.139 & 3079 & 649 \\ 
 ALFALFA & 1.162 & 2.400 & 2.476 & 0.113 & 18713 & 1898 \\ 
 FASHI north & 1.219 & 2.295 & 2.150 & 0.322 & 16879 & 13872 \\ 
 FASHI south & 1.069 & 2.157 & 2.176 & 0.259 & 2720 & 1422 \\ 
 \hline
\end{tabular}
\end{table*}

\subsection{Calculation method for HIMF}
To derive the HIMF, it is also necessary to know the detectable volume for each galaxy. The HIMF is commonly measured with the two-dimensional stepwise maximum likelihood (2DSWML) method \citep[e.g.][]{Efstathiou1988, Zwaan2005, Martin2010, Jones2018} by calculating the effective volume, $V_{\rm eff}$, for each galaxy, which is often referred to as the $1/V_{\rm eff}$ method. The effective volume is obtained by maximising the joint likelihood of finding all sample galaxies in different $M_\hj$ and $W_{50}$ bins and also applied in a non-parametric way (i.e. without assuming a functional form of the HIMF) and in correlation to the galaxy space density. As extensively discussed in \cite{Martin2010}, the $1/V_{\rm eff}$ method has the advantage of being robust against density fluctuations in the large-scale structure, compared to the traditional $1/V_{\rm max}$ method \citep{Schmidt1968}. However, the $1/V_{\rm eff}$ method is very sensitive to the exact completeness cut. As shown in Fig.~1 of \cite{Oman2022}, the HIMF of ALFALFA measured with the $1/V_{\rm eff}$ method is much higher at the low-mass end because it uses a 0.02~dex higher $S_{21,50\%}$ cut from the ALFALFA 100\% sample, compared to the old cut from the 40\% sample. 

On the other hand, the main bias in the $1/V_{\rm max}$ method is the influence of large-scale structures. The ALFALFA volume covers the Virgo Cluster and the Local Supercluster, where the number densities of galaxies with low $M_\hj$ are much higher than those of other regions. The estimated HIMF using $1/V_{\rm max}$ would be systematically overestimated at the low-mass end without corrections \citep{Martin2010}, which motivates the use of the $1/V_{\rm eff}$ method to measure the HIMF for ALFALFA. The application of the $1/V_{\rm eff}$ method is based on the density variation within a given survey volume. Therefore, it still suffers from the effect of large-scale structures when applied to \hi samples in different survey volumes. For example, for the ALFALFA survey, there are large differences between the HIMFs of the spring and fall sky regions derived separately using the $1/V_{\rm eff}$ method \citep[see Fig.~3 of][]{Jones2018}. A homogeneous \hi sample that covers a very large survey volume is ideal for applying the $1/V_{\rm eff}$ method to derive the intrinsic HIMF. 

In this paper, by combining three surveys that cover almost 76\% of the entire sky in the Local Universe, the influence of large-scale structures is effectively suppressed for both the $1/V_{\rm eff}$ and $1/V_{\rm max}$ methods. However, the $1/V_{\rm eff}$ method is not directly applicable in this case because the three surveys have completely different selection functions, as shown in Fig.~\ref{fig:w50s21}. When using the $1/V_{\rm eff}$ method for the combined sample, it would be diffcult to make a fair comparison of the likelihood of finding galaxies in different survey volumes. Therefore, we chose to apply the $1/V_{\rm max}$ method to measure the total HIMF.

The $V_{\rm max}$ value of each galaxy is simply obtained from the maximum distance that the galaxy can be observed with the 50\% completeness limit. Since the HIPASS sample is limited to a slightly smaller redshift of $z<0.042$, we can assume that there is no evolution within the redshift range of $0.042<z<0.05$. The final HIMF can be obtained as,
\begin{equation}
    \phi(M_\hj)=\sum_i\frac{f_{{\rm rms},i}}{C(S_{21,i}|W_{50,i})V_{{\rm max},i}}
    \label{eq:phi}
,\end{equation}
where $C(S_{21,i}|W_{50,i})$ is the completeness of the $i$-th galaxy with a flux density, $S_{21,i}$, and a line profile width $W_{50,i}$. Then, $V_{{\rm max},i}$ is the corresponding maximum volume and $f_{{\rm rms},i}$ is the rms noise correction factor for each galaxy in FASHI calculated in Fig.~\ref{fig:weight}, which we set to 1 for ALFALFA and HIPASS. The sum is over all galaxies in a given $M_\hj$ bin. When combining the four samples, we calculated $V_{\rm max}$ for each galaxy using the total sky area of $31528\,\deg^2$. We note that we only selected galaxies above the 50\% completeness cut and the incompleteness effect is taken into account in Eq.~\ref{eq:phi} with the weight of $1/C(S_{21,i})$ for each galaxy.

\section{H{~\scriptsize I} mass function}\label{sec:results}
\begin{figure*}
\centering
\includegraphics[width=0.9\textwidth]{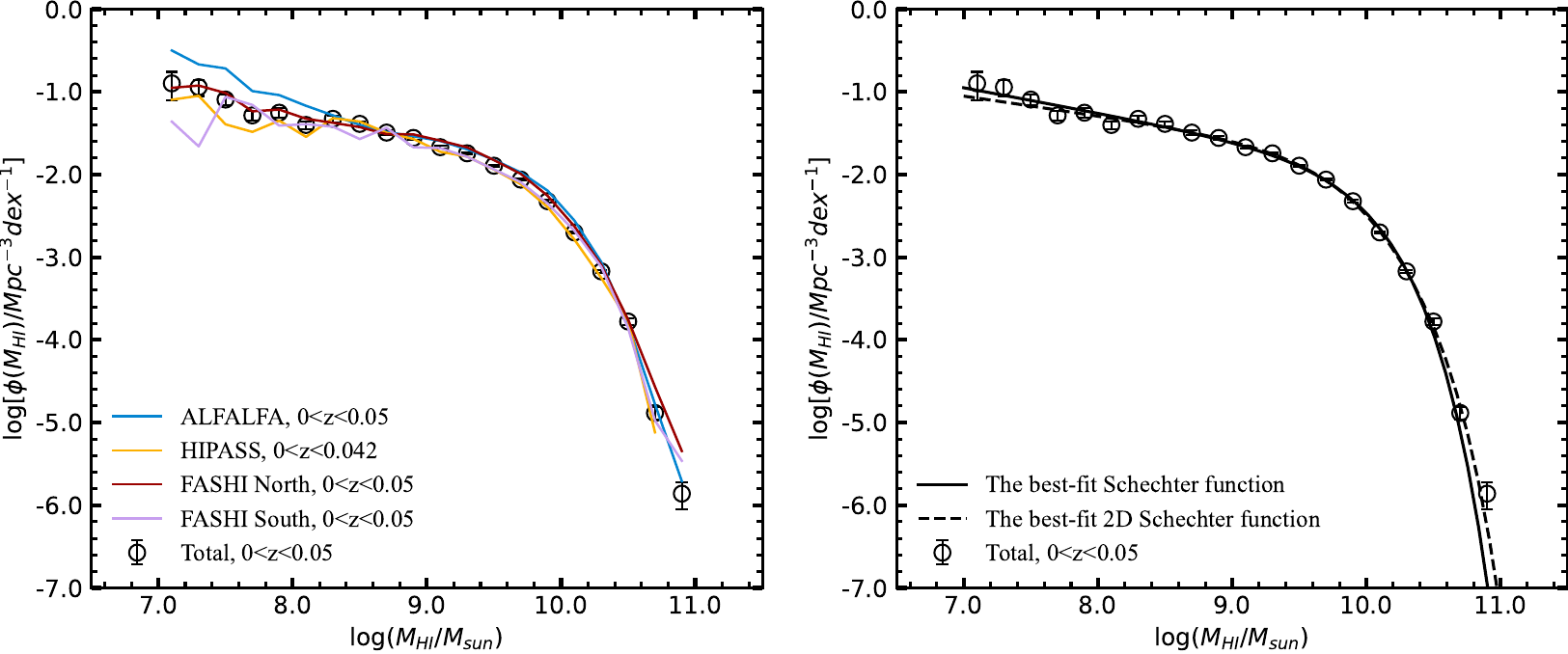}
\caption{Left: \hi mass functions of ALFALFA, HIPASS, and FASHI north and FASHI south, shown as blue, orange, red, and purple solid lines, respectively. The total HIMF by combining three surveys is shown as black open circles with error bars. Right: Best-fit Schechter function (black solid line) and 2D Schechter function (black dashed lines).} 
\label{fig:himf}
\end{figure*}

In the left panel of Fig.~\ref{fig:himf}, we show the total HIMF as open circles, along with our measurements of the HIMFs calculated using the same methods for the individual surveys of ALFALFA, HIPASS, FASHI north, and FASHI south using solid lines of different colours. All measurements are listed in Table~\ref{tab:himf}. The cosmic variance effect is weak for $M_\hj>10^{10}\,\msun$, where the individual HIMFs of different samples are consistent with each other. These galaxies are probed to larger volumes, and thus less affected by the large-scale structures. However, the discrepancies become much larger at the lower-mass end, especially for $M_\hj<10^8\,\msun$. In fact, even for the deepest FASHI north sample, galaxies with $M_\hj<10^8\,\msun$ are only detected within $50\,{\rm Mpc}$, which leads to the large variation of HIMFs at the low-mass end. 

It is interesting that the HIMFs of ALFALFA and FASHI north are consistently higher than those of HIPASS and FASHI south for $M_\hj<10^{10}\,\msun$. This reflects the differences in the large-scale structures of the northern and southern skies. As mentioned above, the superclusters in the northern sky will lead to the overestimate of the HIMF at the low-mass end, whereas the voids presented in the southern sky will cause the underestimate, such as the Local Void (around RA$\sim270^\circ$ and Dec. $\sim-30^\circ$) in the HIPASS footprint \citep{Meyer2004}. Combining the measurements in different surveys of various large-scale structures has significantly improved the estimates of the HIMF in the Local Universe.

For the total HIMF, the measurement errors would be less dominated by the cosmic variance effect, and more so by the Poisson noise caused by the limited numbers of galaxies. Therefore, we can approximate the variance for the total HIMF as in \cite{Jones2018}, namely,
\begin{equation}
    \sigma^2(\phi)=\sum_i\left(\frac{1}{C(S_{21,i})V_{{\rm max},i}}\right)^2.
\end{equation}

Following common practice, we also fit a Schechter function to the total HIMF, shown as the solid line in the right panel of Fig.~\ref{fig:himf}. The best-fitting parameters are $\alpha = -1.30 \pm 0.01$ and $\log (M_s/h_{70}^{-2}\msun) = 9.86 \pm 0.01$ and $\phi_s = (6.58 \pm 0.23) \times 10^{-3}\,h_{70}^3\,{\rm Mpc ^{-3} dex^{-1}}$. The best-fitting parameters for the total HIMF and each of the three surveys are presented in Table~\ref{tab:himf_fit}. However, this single Schechter fitting slightly underestimates the high-mass end of the HIMF (the largest \hi mass bin). It has previously been suggested that a double Schechter function provides a much better fit to the galaxy stellar mass function at low redshifts, especially at the low-mass end \citep[see e.g.][]{Baldry2008,Tomczak2014}. We also adopt the following double-Schechter function to fit the total HIMF, 
\begin{eqnarray}
    \phi(M_{\hj}) &=&
    \ln10\,\phi_{s_1}\left(\frac{M_\hj}{M_{s_1}}\right)^{\alpha+1} \exp\left(-\frac{M_\hj}{M_{s_1}}\right)\nonumber \\
    &+& \ln10\,\phi_{s_2}\left(\frac{M_\hj}{M_{s_2}}\right)^{\alpha+1} \exp\left(-\frac{M_\hj}{M_{s_2}}\right),     
\end{eqnarray}
where the parameters $(\phi_{s_1}, M_{s_1}, \alpha)$ and $(\phi_{s_2}, M_{s_2}, \alpha)$ refer to the two Schechter components with the same slope $\alpha$, respectively. Shown as a dashed line in the right panel of Fig.~\ref{fig:himf}, it provides a better fit than a single Schechter function, especially at the high-mass end. We find that the total HIMF is better fitted with the same slope, $\alpha$, and two different `knee' masses, $M_{s_1}$ and $M_{s_2}$. The best-fitting reduced $\chi^2$ (i.e. $\chi^2/{\rm dof}$) decreases from $56/17$ for the single-Schechter function to $24/15$ for the double-Schechter function. The best-fitting parameters are $\phi_{s_1}= (2.67 \pm 0.98)\times 10^{-3}\,h_{70}^3\,{\rm Mpc ^{-3}dex^{-1}}$, $\log(M_{s_1}/h_{70}^{-2}\msun) = 9.96 \pm 0.03$, $\phi_{s_2}= (5.96 \pm 0.78)\times 10^{-3}\,h_{70}^3\,{\rm Mpc^{-3} dex^{-1}}$, $\log(M_{s_2}/h_{70}^{-2}\msun) = 9.65 \pm 0.07$, and $\alpha = -1.24 \pm 0.02$. The two different `knee' masses are likely associated with different galaxy populations, for example, central and satellite galaxies. It will be further confirmed in our future work with the decomposition of HIMF into central and satellite galaxies.

The cosmic \hi abundance, $\Omega_{\hj}$, can be estimated by integrating the best-fitting single Schechter function as follows \citep{Martin2010,Jones2018}, 
\begin{equation}
    \Omega_{\hj} = \frac{1}{\rho_c}\phi_s M_s \Gamma(\alpha+2),
\end{equation}
where $\rho_c$ is the critical density at $z=0$ (we assume $H_0 = 70 \,h_{70}\,\kms \, {\rm Mpc}^{-1}$). This gives $\Omega_{\hj} = (4.54 \pm 0.20)\times 10^{-4}\,h_{70}^{-1}$, which is almost the same for double Schechter function fits. 

\begin{table*}
\setlength{\tabcolsep}{3pt} 
\renewcommand{\arraystretch}{1.3} 
\caption{HIMFs of the three surveys and the total results of their combination.}\label{tab:himf}
\centering
\begin{tabular}{ c c c c c c } 
 \hline
 \hline
 $\log (M_\hj/\msun)$ & \multicolumn{5}{c}{$\phi(M_\hj)$ with errors ($\rm Mpc^{-3}dex^{-1}$)}   \\ 
 & HIPASS & ALFALFA & FASHI north & FASHI south & Total \\
 \hline
 7.1 & { \small$( 8.036 \pm 8.036 ) \times 10^{-2}$ } & { \small $(3.163 \pm 0.537 ) \times 10^{-1}$ } & { \small $(1.118\pm 0.255 ) \times 10^{-1}$ } & { \small $(4.363\pm 4.372 ) \times 10^{-2}$ } & { \small $(1.266\pm 0.479 ) \times 10^{-1}$ } \\  
 7.3 & { \small$( 8.957 \pm 4.110 ) \times 10^{-2}$ } & { \small $(2.147 \pm 0.350 ) \times 10^{-1}$ } & { \small $(1.191\pm 0.204 ) \times 10^{-1}$ } & { \small $(2.192\pm 1.676 ) \times 10^{-2}$ } & { \small $(1.140\pm 0.250 ) \times 10^{-1}$ } \\  
 7.5 & { \small$( 4.045 \pm 1.852 ) \times 10^{-2}$ } & { \small $(1.919 \pm 0.250 ) \times 10^{-1}$ } & { \small $(9.678\pm 1.408 ) \times 10^{-2}$ } & { \small $(8.694\pm 2.899 ) \times 10^{-2}$ } & { \small $(8.063\pm 1.204 ) \times 10^{-2}$ } \\  
 7.7 & { \small$( 3.278 \pm 1.108 ) \times 10^{-2}$ } & { \small $(1.022 \pm 0.151 ) \times 10^{-1}$ } & { \small $(5.811\pm 0.758 ) \times 10^{-2}$ } & { \small $(6.954\pm 1.831 ) \times 10^{-2}$ } & { \small $(5.191\pm 0.718 ) \times 10^{-2}$ } \\  
 7.9 & { \small$( 4.486 \pm 1.172 ) \times 10^{-2}$ } & { \small $(9.135 \pm 0.873 ) \times 10^{-2}$ } & { \small $(6.103\pm 0.595 ) \times 10^{-2}$ } & { \small $(3.903\pm 1.787 ) \times 10^{-2}$ } & { \small $(5.593\pm 0.713 ) \times 10^{-2}$ } \\  
 8.1 & { \small$( 2.857 \pm 0.659 ) \times 10^{-2}$ } & { \small $(6.782 \pm 0.544 ) \times 10^{-2}$ } & { \small $(4.724\pm 0.388 ) \times 10^{-2}$ } & { \small $(4.067\pm 0.860 ) \times 10^{-2}$ } & { \small $(3.975\pm 0.404 ) \times 10^{-2}$ } \\  
 8.3 & { \small$( 4.807 \pm 0.695 ) \times 10^{-2}$ } & { \small $(5.202 \pm 0.349 ) \times 10^{-2}$ } & { \small $(4.227\pm 0.285 ) \times 10^{-2}$ } & { \small $(3.796\pm 0.774 ) \times 10^{-2}$ } & { \small $(4.716\pm 0.413 ) \times 10^{-2}$ } \\  
 8.5 & { \small$( 4.373 \pm 0.513 ) \times 10^{-2}$ } & { \small $(3.989 \pm 0.226 ) \times 10^{-2}$ } & { \small $(3.746\pm 0.227 ) \times 10^{-2}$ } & { \small $(2.687\pm 0.380 ) \times 10^{-2}$ } & { \small $(4.098\pm 0.304 ) \times 10^{-2}$ } \\  
 8.7 & { \small$( 3.203 \pm 0.295 ) \times 10^{-2}$ } & { \small $(3.275 \pm 0.150 ) \times 10^{-2}$ } & { \small $(3.078\pm 0.143 ) \times 10^{-2}$ } & { \small $(3.667\pm 0.398 ) \times 10^{-2}$ } & { \small $(3.217\pm 0.177 ) \times 10^{-2}$ } \\  
 8.9 & { \small$( 2.695 \pm 0.212 ) \times 10^{-2}$ } & { \small $(2.899 \pm 0.106 ) \times 10^{-2}$ } & { \small $(3.057\pm 0.109 ) \times 10^{-2}$ } & { \small $(2.139\pm 0.197 ) \times 10^{-2}$ } & { \small $(2.771\pm 0.127 ) \times 10^{-2}$ } \\  
 9.1 & { \small$( 1.884 \pm 0.131 ) \times 10^{-2}$ } & { \small $(2.547 \pm 0.075 ) \times 10^{-2}$ } & { \small $(2.569\pm 0.078 ) \times 10^{-2}$ } & { \small $(2.085\pm 0.167 ) \times 10^{-2}$ } & { \small $(2.142\pm 0.079 ) \times 10^{-2}$ } \\  
 9.3 & { \small$( 1.650 \pm 0.097 ) \times 10^{-2}$ } & { \small $(2.019 \pm 0.050 ) \times 10^{-2}$ } & { \small $(2.129\pm 0.054 ) \times 10^{-2}$ } & { \small $(1.641\pm 0.102 ) \times 10^{-2}$ } & { \small $(1.806\pm 0.058 ) \times 10^{-2}$ } \\  
 9.5 & { \small$( 1.149 \pm 0.064 ) \times 10^{-2}$ } & { \small $(1.510 \pm 0.032 ) \times 10^{-2}$ } & { \small $(1.494\pm 0.034 ) \times 10^{-2}$ } & { \small $(1.155\pm 0.069 ) \times 10^{-2}$ } & { \small $(1.279\pm 0.038 ) \times 10^{-2}$ } \\  
 9.7 & { \small$( 7.646 \pm 0.395 ) \times 10^{-3}$ } & { \small $(1.060 \pm 0.020 ) \times 10^{-2}$ } & { \small $(1.020\pm 0.022 ) \times 10^{-2}$ } & { \small $(8.062\pm 0.463 ) \times 10^{-3}$ } & { \small $(8.677\pm 0.237 ) \times 10^{-3}$ } \\  
 9.9 & { \small$( 4.032 \pm 0.250 ) \times 10^{-3}$ } & { \small $(6.383 \pm 0.124 ) \times 10^{-3}$ } & { \small $(5.525\pm 0.130 ) \times 10^{-3}$ } & { \small $(4.463\pm 0.251 ) \times 10^{-3}$ } & { \small $(4.757\pm 0.150 ) \times 10^{-3}$ } \\  
 10.1 & { \small$( 1.629 \pm 0.117 ) \times 10^{-3}$ } & { \small $(2.773 \pm 0.064 ) \times 10^{-3}$ } & { \small $(2.358\pm 0.067 ) \times 10^{-3}$ } & { \small $(2.085\pm 0.144 ) \times 10^{-3}$ } & { \small $(1.995\pm 0.070 ) \times 10^{-3}$ } \\  
 10.3 & { \small$( 5.612 \pm 0.524 ) \times 10^{-4}$ } & { \small $(8.753 \pm 0.315 ) \times 10^{-4}$ } & { \small $(8.044\pm 0.327 ) \times 10^{-4}$ } & { \small $(7.612\pm 0.670 ) \times 10^{-4}$ } & { \small $(6.737\pm 0.317 ) \times 10^{-4}$ } \\  
 10.5 & { \small$( 1.703 \pm 0.240 ) \times 10^{-4}$ } & { \small $(1.543 \pm 0.125 ) \times 10^{-4}$ } & { \small $(1.754\pm 0.133 ) \times 10^{-4}$ } & { \small $(1.386\pm 0.230 ) \times 10^{-4}$ } & { \small $(1.667\pm 0.144 ) \times 10^{-4}$ } \\  
 10.7 & { \small$( 7.614 \pm 3.232 ) \times 10^{-6}$ } & { \small $(1.670 \pm 0.405 ) \times 10^{-5}$ } & { \small $(2.709\pm 0.506 ) \times 10^{-5}$ } & { \small $(1.055\pm 0.611 ) \times 10^{-5}$ } & { \small $(1.307\pm 0.225 ) \times 10^{-5}$ } \\  
 10.9 & - & { \small $(1.951 \pm 1.380 ) \times 10^{-6}$ } & { \small $(4.523\pm 2.095 ) \times 10^{-6}$ } & { \small $(3.448\pm 3.434 ) \times 10^{-6}$ } & { \small $(1.382\pm 0.499 ) \times 10^{-6}$ } \\ 
 \hline
\end{tabular}
\end{table*}

\begin{table*}
\setlength{\tabcolsep}{15pt} 
\renewcommand{\arraystretch}{1.3} 
\caption{HIMF fitting parameters for the three surveys and the total results of their combination.}\label{tab:himf_fit}
\centering
\begin{tabular}{ c c c c  } 
 \hline
 \hline
 Survey & $\alpha$ & $\log(M_s/h_{70}^{-2}\msun)$  &  $\phi_s/h_{70}^3 {\rm Mpc}^{-3} {\rm dex}^{-1}$ \\ 
 \hline
 Total & $-1.30 \pm 0.01$ & $9.86 \pm 0.01$ & $(6.58 \pm 0.23)\times10^{-3}$ \\
 HIPASS & $-1.28 \pm 0.02$ & $ 9.83 \pm 0.02$ & $(6.37 \pm 0.42)\times10^{-3}$ \\
 ALFALFA & $-1.29 \pm 0.01$ & $9.87 \pm 0.01$ & $(7.91 \pm 0.22)\times10^{-3}$ \\
 FASHI north & $-1.26 \pm 0.01$ & $9.84 \pm 0.01$ & $(8.26 \pm 0.24)\times10^{-3}$ \\
 FASHI south & $-1.25 \pm 0.03$ & $9.86 \pm 0.02$ & $(6.36 \pm 0.43)\times10^{-3}$ \\ 
 \hline
\end{tabular}
\end{table*}

\begin{figure}
\centering
\includegraphics[width=\columnwidth]{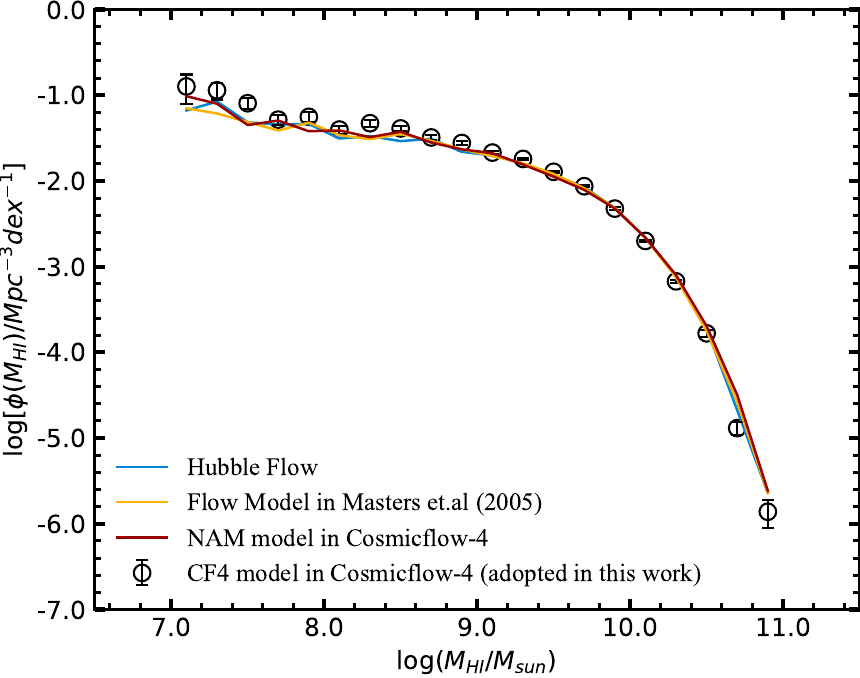}
\caption{Total \hi mass functions calculated by different distance estimate models. The pure Hubble Flow, flow model in \cite{Masters2005}, NAM model and CF4 model in Cosmicflow-4 (adopted in our work) are shown as blue, orange, red solid lines, and black open circles with their errorbars, respectively.} 
\label{fig:himf_dis}
\end{figure}

\section{Discussion}\label{sec:discussion}
\subsection{Distance uncertainties}\label{subsec:distance_error}

As discussed in previous sections, measurement errors in distance estimates could potentially lead to systematic uncertainties in the HIMF \citep{Masters2004}. To investigate the effects of distance estimates, we followed the practice of \cite{Jones2018} to measure the HIMFs by applying different flow models. In Fig.~\ref{fig:himf_dis}, we compare the total HIMFs calculated using different distance models, including the pure Hubble flow (blue line: assuming $H_0 = 70\,h_{70}\kms\,{\rm Mpc}^{-1}$), the flow model of \citet[orange line]{Masters2005}: the NAM model (red line), and the CF4 model (adopted in this work) in the Cosmicflows-4 Distance–Velocity Calculator. The distance estimates in the NAM model are limited to 38 Mpc, whereas the CF4 model extends this limit to 500 Mpc. The flow model of \cite{Masters2005} is also limited to a distance of $cz_\odot<6000\,\kms$ as in the ALFALFA sample. For galaxies lying beyond these distances, we utilised the pure Hubble flow. 

The HIMF with the CF4 model is consistent with all other models for $M_\hj>10^9\,\msun$, but is slightly higher at the low-mass end. As shown in \cite{Tully2023}, by including the kinematic information of ALFALFA, the peculiar velocities in Cosmicflows-4 are much improved compared to Cosmicflows-3. The maximum difference in these distance estimates for $M_\hj<10^8\,\msun$ is around $\Delta\log D_L\sim0.08$, which will introduce an offset of 0.16~dex in the \hi mass at the low-mass end. Since the HIMF at the low-mass end is $\phi(M_\hj)\propto M_\hj^{\alpha+1}$, the 0.16~dex offset in $M_\hj$ will translate to a minor offset of $0.048$~dex in $\phi(M_\hj)$ for $\alpha=-1.30$. The small number of galaxies at the low-mass end also limits our ability to tightly constrain the HIMF here. For the massive end, the effects of different flow models are weaker with respect to the pure Hubble flow. Therefore, we can conclude that the effect of different distance estimates would not substantially change the measured HIMF, consistent with the results shown in \cite{Jones2018}. 

\subsection{Comparison with the literature}

\begin{figure}
\centering
\includegraphics[width=\columnwidth]{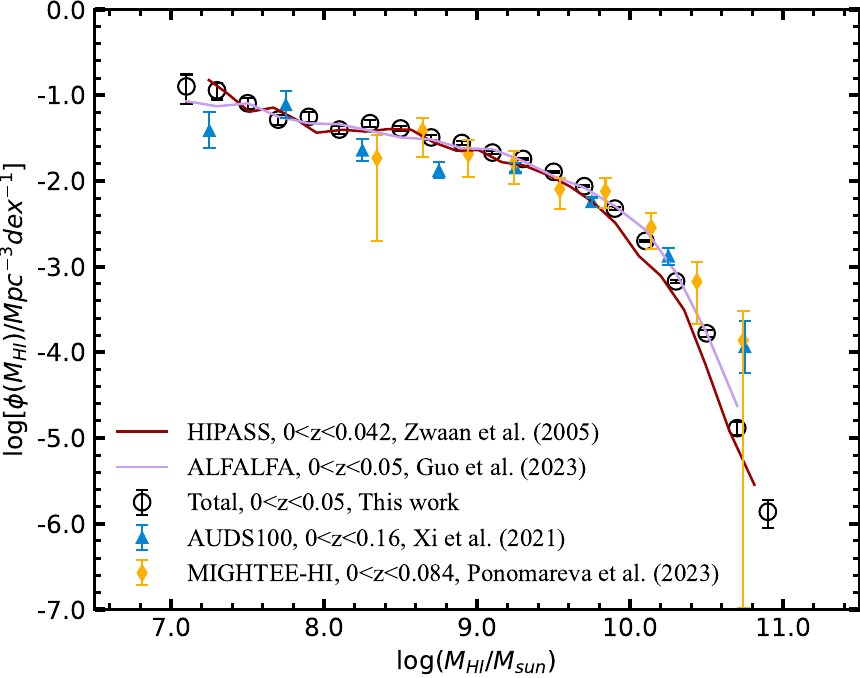}
\caption{Comparisons of HIMF in a range of literature works. Measurements of HIPASS in \cite{Zwaan2005}, ALFALFA in \cite{Guo2023}, AUDS100 in \cite{Xi2021}, and MIGHTEE--HI in \cite{Ponomareva2023} are shown as the red solid line, purple solid line, blue triangles, and yellow diamonds, respectively. The total HIMF derived in this work are shown as black open circles with error bars.} 
\label{fig:com_himf}
\end{figure}

In Fig.~\ref{fig:com_himf}, we compare our HIMF measurement (open circles) with those of HIPASS from \cite{Zwaan2005}, ALFALFA from \cite{Guo2023},  Arecibo Ultra-Deep Survey (AUDS) from \cite{Xi2021}, and MIGHTEE-HI from \cite{Ponomareva2023}. 

The HIMF of HIPASS in \cite{Zwaan2005} shows a slightly lower high-mass end amplitude. We note that their original HIMF measurements used a higher Hubble constant of $H_0=75\,h_{75}\kms\,{\rm Mpc}^{-1}$. For consistency with our definition of $H_0=70\,h_{70}\kms\,{\rm Mpc}^{-1}$, we applied approximate corrections to increase their $M_\hj$ by 0.06~dex and decreasing $\phi(M_\hj)$ by $-0.09$~dex. The underestimation of $\phi(M_\hj)$ at the high-mass end could then be caused by uncertainties in the completeness function and the HIMF normalisation estimation method \citep{Zwaan2003}. They adopted a different completeness estimation method using the recovered rates of the inserted synthetic sources \citep{Zwaan2004}. In their 2DSWML method, the overall normalisation of HIMF is lost in the maximum likelihood estimator and is then determined from the mean galaxy number density with the minimum variance estimator \citep{Zwaan2005}.  

The HIMF measurement of \cite{Guo2023} used the ALFALFA 100\% sample and corrected for the incompleteness effect in \cite{Jones2018} by using the 90\% completeness cut $S_{21,90\%}$ of \cite{Haynes2011}. Their HIMF and $\Omega_\hj= (4.55 \pm 0.29) \times 10^{-4}\,h_{70}^{-1}$ results are quite consistent with ours. We also show the AUDS measurement of \cite{Xi2021} that extends to a slightly higher redshift of $z=0.16$. It shows a mild redshift evolution of the HIMF, with a shallower slope at the high-mass end. However, their \hi detection only included 247 galaxies in a sky area of $1.35\deg^2$. The small number of galaxies limits an accurate determination of the HIMF. However, they found $\Omega_\hj=(3.93\pm0.68)\times 10^{-4}\,h_{70}^{-1}$, which is consistent with our $\Omega_\hj$ measurement. \cite{Ponomareva2023} measured the HIMF using the MIGHTEE-HI survey in a redshift range of $0\le z\le0.084$. They also found slightly higher $\phi(M_\hj)$ amplitudes in $M_\hj>10^{10}\,\msun$ than in the $z=0$ measurements. Their $\Omega_\hj$ measurement is slightly larger, with $\Omega_\hj=5.46^{+0.94}_{-0.99}\times10^{-4}\,h_{67.4}^{-1}$, albeit with large errors. These measurements tend to point to the weak evolution of $\Omega_\hj$ in $0<z<0.2$ \citep{Rhee2018,Walter2020}, consistent with the prediction of the theoretical model \citep{Guo2023}. Ongoing and future deep \hi observations, for example, the FAST Ultra-Deep Survey \citep{Xi2024}, will provide more insight into the evolution of the HIMF.

We note that the HIMF of our total sample is consistent with the HIMF of \cite{Guo2023}. However, from the best-fitting Schechter function fits displayed in Table~\ref{tab:himf_fit}, the relevant fitting parameters are slightly different. We have a slightly lower `knee' mass of $\log(M_s/h_{70}^{-2}\msun) = 9.86 \pm 0.01$, while it is reported as $\log(M_s/h_{70}^{-2}\msun) = 9.91 \pm 0.01$ in \cite{Guo2023}. However, as shown in Fig.~4 of \cite{Ponomareva2023}, the three parameters of the Schechter function are strongly correlated with each other, with $M_s$ showing anticorrelations with $\alpha$ and $\phi_s$. Comparisons between the parameters of Schechter function fits of different samples should not be treated independently.

\section{Conclusions}\label{sec:conclusion}
In this study, we measured the \hi mass function in the Local Universe ($0<z<0.05$) by combining the \hi samples in the HIPASS, ALFALFA and FASHI surveys, covering 76\% of the entire sky ($31528\,\deg^2$). The combined sample has the advantage of greatly suppressing the influence of cosmic variance on the measured HIMF. To further reduce systematic uncertainties in the processing of the \hi catalogues, we adopted the same methods to estimate distances, calculate sample completeness, and determine the HIMF (the $1/V_{\rm max}$ method) for all three catalogues. We measured the most complete HIMF in the Local Universe so far (Fig.~\ref{fig:himf}) and  obtained the first HIMF measurement for the recent FASHI survey, presented here.  

We fit the total HIMF with a single Schechter function, with the parameters of $\alpha = -1.30 \pm 0.01$ and $\log (M_s/h_{70}^{-2}\msun) = 9.86 \pm 0.01$ and $\phi_s = (6.58 \pm 0.23) \times 10^{-3}\,h_{70}^3\,{\rm Mpc ^{-3} dex^{-1}}$. The derived cosmic \hi abundance is $\Omega_{\hj} = (4.54 \pm 0.20)\times 10^{-4}\,h_{70}^{-1}$, which is consistent with the measurement using the ALFALFA 100\% complete sample \citep{Guo2023}. However, we find that our HIMF is better described by a double Schechter function with the same slope $\alpha$. The best-fitting parameters are $\phi_{s_1}= (2.67 \pm 0.98)\times 10^{-3}\,h_{70}^3\,{\rm Mpc ^{-3} dex^{-1}}$, $\log(M_{s_1}/h_{70}^{-2}\,\msun) = 9.96 \pm 0.03$, $\phi_{s_2}= (5.96 \pm 0.78)\times 10^{-3}\,h_{70}^3\,{\rm Mpc ^{-3} dex^{-1}}$, $\log(M_{s_2}/h_{70}^{-2}\msun) = 9.65 \pm 0.07$, and $\alpha = -1.24 \pm 0.02$. The two different `knee' masses are favoured by the measured HIMF at the massive end, indicating contributions from two different components (likely from the central and satellite galaxies).

We found that the measured HIMF is marginally affected by the choice of distance estimates. We adopted different flow models to estimate the luminosity distances and obtained fully consistent results. However, local large-scale structures have a strong influence on the HIMF when measured separately in different \hi samples, especially at the low-mass end. ALFALFA and FASHI north have consistently higher HIMFs than those of HIPASS and FASHI south, due to the influence of local superclusters. Combining the three \hi surveys provides a unique opportunity to obtain an unbiased estimate of the HIMF in the Local Universe. 

We note that although the combined sample covers a large sky area, galaxies with low $M_\hj$ have been  probed within very limited volumes and with limited statistics. Deeper \hi surveys in the near future will provide more robust measurements of HIMF at the low-mass end. 
        
\begin{acknowledgements}
We thank the anonymous reviewer for the helpful comments that improved the presentation of this paper. This work is supported by the National SKA Program of China (grant No. 2020SKA0110100), the Guizhou Provincial Science and Technology Projects (QKHFQ[2023]003, QKHPTRC-ZDSYS[2023]003, QKHFQ[2024]001-1, QKHJC-ZK[2025]MS015) and the CAS Project for Young Scientists in Basic Research (No. YSBR-092). We acknowledge the use of the High Performance Computing Resource in the Core Facility for Advanced Research Computing at the Shanghai Astronomical Observatory.
\end{acknowledgements}

\bibliographystyle{aa}
\bibliography{ref}
        
\end{document}